\documentclass[preprintnumbers, floatfix, preprintnumbers, letterpaper, superscriptaddress,nofootinbib, twocolumn]{revtex4}
\pdfoutput=1
\usepackage{graphicx}
\usepackage{microtype}
\usepackage{amsmath}
\usepackage{amssymb}
\usepackage{subfigure}
\usepackage{hyperref}
\usepackage{url}
\usepackage{xcolor}
\usepackage{color}
\usepackage{mathrsfs}
\usepackage{calrsfs}
\usepackage{amsfonts}
\usepackage{latexsym}
\usepackage{ragged2e}
\usepackage{epsfig}
\usepackage{textcomp}
\usepackage{phaistos}
\makeatletter
\renewcommand\@makefnmark{\hbox{\@textsuperscript{\normalfont\color{purple}\@thefnmark}}}
\renewcommand\@makefntext[1]{%
  \parindent 1em\noindent
            \hb@xt@1.8em{%
                \hss\@textsuperscript{\normalfont\@thefnmark}}#1}
\makeatother

\usepackage{caption}
\DeclareCaptionJustification{justified}{\leftskip=0pt \rightskip=0pt \parfillskip=0pt plus 1fil}
\definecolor{vividviolet}{rgb}{0.62, 0.0, 1.0}
\definecolor{amaranth}{rgb}{0.9, 0.17, 0.31}
\definecolor{palatinateblue}{rgb}{0.15, 0.23, 0.89}
\definecolor{brightpink}{rgb}{1.0, 0.0, 0.5}
\definecolor{cornflowerblue}{rgb}{0.39, 0.58, 0.93}
\definecolor{deepcarminepink}{rgb}{0.94, 0.19, 0.22}
\definecolor{radicalred}{rgb}{1.0, 0.21, 0.37}

\hypersetup{ linktoc=all,
    colorlinks, linkcolor={palatinateblue},
    citecolor={brightpink}, urlcolor={amaranth}
}

\graphicspath{{Images/}}

\renewcommand{\d}[1]{\ensuremath{\operatorname{d}\!{#1}}}

\graphicspath{{Images/}}
\renewcommand{\d}[1]{\ensuremath{\operatorname{d}\!{#1}}}
\makeatletter
\def\@fnsymbol#1{\ensuremath{\ifcase#1\or $\PHplaneTree$ \or $\textleaf$
\else\@ctrerr\fi}}%

\makeatother

\def\sideremark#1{\ifvmode\leavevmode\fi\vadjust{\vbox to0pt{\vss
 \hbox to 0pt{\hskip\hsize\hskip1em
 \vbox{\hsize1.5cm\tiny\raggedright\pretolerance10000
 \noindent #1\hfill}\hss}\vbox to8pt{\vfil}\vss}}}%
\begin{document}

\title{Cosmic Censorship and the Evolution of $d$-Dimensional \\Charged Evaporating Black Holes}

\author{Hao Xu}
\email{xuh5@sustech.edu.cn}
\affiliation{Institute for Quantum Science and Engineering, Department of Physics, Southern University of Science and Technology, Shenzhen 518055, China}
\affiliation{Department of Physics, University of Science and Technology of China, Hefei 230026, China}

\author{Yen Chin \surname{Ong}}
\email{ycong@yzu.edu.cn}
\affiliation{Center for Gravitation and Cosmology, College of Physical Science and Technology, Yangzhou University, \\180 Siwangting Road, Yangzhou City, Jiangsu Province  225002, China}
\affiliation{School of Aeronautics and Astronautics, Shanghai Jiao Tong University, Shanghai 200240, China}

\author{Man-Hong Yung}
\affiliation{Institute for Quantum Science and Engineering, Department of Physics, Southern University of Science and Technology, Shenzhen 518055, China}
\affiliation{Shenzhen Key Laboratory of Quantum Science and Engineering, Southern University of Science and Technology Shenzhen, 518055, China}

\begin{abstract}
The cosmic censorship conjecture essentially states that naked singularities should not form from \emph{generic} initial conditions.
Since black hole parameters can change their values under Hawking evaporation, one has to ask whether it is possible to reach extremality by simply waiting for the black hole to evaporate. If so a slight perturbation would likely render the singularity naked.
Fortunately, at least for the case of asymptotically flat 4-dimensional Reissner-Nordstr\"om black hole,
Hiscock and Weems showed that it can never reach extremality despite the fact that for a sufficiently massive black hole, its charge-to-mass ratio can increase during Hawking evaporation. Hence cosmic censorship is never violated by Hawking emission.
However, we know that under some processes, it is easier to violate cosmic censorship in higher dimensions, therefore it is crucial to generalize Hiscock and Weems model to dimensions above four to check cosmic censorship. We found that Hawking evaporation cannot lead to violation of cosmic censorship even in higher dimensional Reissner-Nordstr\"om spacetimes. Morerover, it seems to be more difficult to reach extremality as number of dimension increases.
\end{abstract}

\maketitle

\section{Introduction: Hawking Evaporation and Cosmic Censorship}

Black holes hide singularities behind their horizons.
Singularities themselves are problematic in general relativity, and if it ever becomes naked would render general relativity indeterministic. This is because one has to prescribe boundary conditions on the singularities, which we cannot do without understanding what they are.
Nowadays we know that naked singularities can form under various circumstances \cite{1006.5960,1512.04532,1906.08265,1702.01755,1906.10696,9301052,9404071,1107.5821,1812.05017}, notwithstanding the (weak) cosmic censorship conjecture, which states that naked singularities should not form from generic initial conditions \cite{1,1-2}.  The strong cosmic censorship goes a step further, demanding that determinism should hold even inside black holes (in practice it requires the inner horizon of a black hole to be unstable).

One obvious problem is to properly understand what ``generic'' means in this context, and that involves the study of different mechanisms that could lead to naked singularities.
It would be helpful for future research if the community can classify under what physical processes can the censorship be violated.
Since Hawking radiation can change the values of the black hole parameters (i.e. the black hole hairs. In genereal relativity this means mass, charge, and angular momentum), one such mechanism that should be considered is the Hawking evaporation.

Let us consider an isolated\footnote{This assumption ensures that the black hole charge is not neutralized by accreting oppositely charged particles from the surrounding environment. In astrophysical context it has always been assumed that charge plays minimal, if any, role. However, this might be an over-simplification, since there exists some mechanism that could ``charge-up'' a black hole in astrophysics (e.g. the Wald mechanism \cite{wald}), and although the amount of charges do not significantly affect the near-Schwarzschild spacetime geometry, it does affect the motion of charged particles enough to cause degeneracy with effect from rotation \cite{1904.04654}. {In addition, there exist dark matter models (e.g. minicharged dark matter) in which astrophysical black holes might be charged under a hidden U(1) symmetry, and the metric is formally the same as Kerr-Newman solution \cite{1604.07845}.}} asymptotically flat Reissner-Nordstr\"om black hole. {The Hawking evaporation process of this type of black hole has been investigated by many previous works, e.g., \cite{zaumen, gibbons, page1976, 0103090}, just to name a few.

Surprisingly, the charge-to-mass ratio $Q/M$ of  Reissner-Nordstr\"om black holes is not a monotonically decreasing function of time as the black holes undergo Hawking evaporation.} As shown in the work of Hiscock and Weems \cite{HW}, depending on the initial conditions, $Q/M$ can actually \emph{increase} during the course of its evaporation. Sufficiently large black holes can get very close to extremality (for Reissner-Nordstr\"om black holes, extremality is characterized by zero temperature), \emph{even if} it initially started out with a very small charge-to-mass ratio. Suppose that these black holes can indeed become extremal under Hawking evaporation, then they could very well under slight perturbation become naked singularities. Of course, reaching extremality in finite time is itself a violation of the third law of black hole thermodynamics. Indeed, Davies realized a long time ago that ``unattainability of absolute zero (temperature) is equivalent to the cosmic censorship hypothesis'' \cite{davies1977}. This would spell disaster for the cosmic censorship conjecture -- for any practical meaning of genericity --  since a large class of black holes can evolve towards naked singularity end state just by waiting long enough for Hawking radiation to take effect.

Fortunately, this does not occur, at least in the case of Reissner-Nordstr\"om black holes. As shown by Hiscok and Weems, although the charge-to-mass ratio $Q/M$ might come very close to achieve extremality, it never does. Instead, the evolution would eventually turn around towards the Schwarzschild state. More specifically,
 on the Reissner-Nordstrom configuration space of $({Q}/{M})^2$ plotted against $M$, the black hole evolution curves split it into two parts separated by a narrow region which is a dissipative attractor. The attractor region approaches the Schwarzschild limit $({Q}/{M})^2 =0$ when $M\rightarrow 0$, and approaches the extreme Reissner-Nordstrom limit $({Q}/{M})^2=1$ when $M\rightarrow \infty$. If the initial black hole is on the right of the attractor (mass dissipation zone), it will lose mass while the charge is almost unchanged -- thereby increasing $Q/M$ -- until it ``hits'' the attractor. Then the mass and charge continue to decrease and evolve towards the Schwarzschild limit. If the initial black hole is on the left of the attractor (charge dissipation zone), it will lose charge faster than mass, and $Q/M$ will monotonically decrease until it joins the attractor and evolves towards the Schwarzschild limit. That is to say, the attractor is approximately characterized by $\d M/\d t = \d Q/\d t$ \cite{HW,1909.09981}.

Recently, the Hiscock-Weems model has been applied to asymptotically flat dilatonic charged black holes \cite{1907.07490}. Remarkably, if we \emph{impose} the validity of the weak cosmic censorship, then this would force us to modify the charge loss rate, the result of which agrees with direct calculation of charge loss rate via the WKB (Wentze-Kramers-Brillouin) method which considers the scattering of charged scalar on the curved black hole background \cite{1305.2564}. This strongly suggests that cosmic censorship does indeed hold under Hawking evaporation. (It also serves as an example that cosmic censorship can sometimes be used as a useful principle to deduce some other physics.)  In addition, a similar technique was recently applied to study charged black holes in a de Sitter universe \cite{1910.01648}.

Surprisingly, the study of charged black holes evaporation in higher dimensions using the method of Hiscock and Weems is still lacking. In view of its connection to cosmic censorship conjecture, this is a well-motivated problem. After all, we know from the literature that whether naked singularities can form from various processes does depend on the number of spacetime dimensions. For example, some of the earliest example of naked singularity formation involves pinching-off of black strings (Gregory-Laflamme instability \cite{9301052, 9404071, 1906.10696,1107.5821}) which takes place in higher dimensions. More recently, it was discovered that collisions of black holes can violate cosmic censorships in higher dimensions as energy that gets dissipated in gravitational waves decreases when dimensionality increases  \cite{1812.05017}; {see also the earlier result of \cite{1105.3331}, and its recent follow-up \cite{1909.02997}.} (For a short readable account of the maximal amount of gravitational radiation emitted by black hole collision in 4-dimensions, see Hawking's paper \cite{prl1971}.)

Nevertheless, it is not always the case that higher dimension leads to easier violation of cosmic censorship. Recently, strong cosmic censorship of charged black holes in asymptotically de Sitter spacetime has attracted a lot of interest \cite{1711.10502,1808.03631,1811.08538}. The situation in higher dimensions is not so straightforward \cite{1902.01865}: it is indeed easier to violate strong cosmic censorship in higher dimensions, provided that the black holes are ``large'' (in the sense of large $\Lambda/\Lambda_\text{max}$, where $\Lambda$ is the cosmological constant, and $\Lambda_\text{max}$ is the value for which if $\Lambda > \Lambda_\text{max}$, then the spacetime admits at most one horizon). If the black holes are small, then it becomes harder to violate the censorship. Actually the situation is even more complicated: there exists range of $\Lambda/\Lambda_\text{max}$ for which the difficulty to violate strong cosmic censorship is not monotonic as one increases the dimensionality. For example, it could be that it is easiest to achieve this in 6-dimensions, but hardest to do so in 4-dimensions, and \emph{then} followed by 5-dimensions (see Table 1 of \cite{1902.01865}).
In short, it is not easy to tell if under a given process, whether going to higher dimensions makes it easier to violate cosmic censorship (of either version).
Therefore, although Reissner-Nordstr\"om black holes cannot attain extremality in 4-dimensions, one has to ask whether this is still true in higher dimensions, and if so, whether the tendency to achieve extremality, in some suitable measure, is higher or lower.

Our paper is organized as follows. In the next section we will derive the differential equations describing the evolution of charged black hole as functions of time in arbitrary dimensions above four. In Sect.(III) we present some numerical examples of the evaporative evolution of charged black holes. In the final section we give some concluding remarks. We shall adopt the same units of Hiscock and Weems (hereinafter, HW), setting the speed of light in vacuum $c$, the gravitational constant $G$, and the Boltzmann constant $k$ equal to one. However the reduced Planck constant $\hbar$ is retained explicitly.

\section{Hiscock and Weems Model in $d$-Dimensions}

For a  $d$-dimensional asymptotically flat spacetime with an electric field, the action can be written as
\begin{equation}
S = \frac{1}{16{\pi}G} \int_{M} \text{d}^{d}x \sqrt{-g} \left[R - F^2 \right],
\label{action}
\end{equation}
where $R$ is the Ricci scalar and $F^2$ the contraction of the Maxwell tensor $F^2:=F^{\mu \nu}F_{\mu \nu} $.
The asymptotically flat $d$-dimensional ($d \geqslant$ 4) charged black hole (Reissner-Nordstr\"om solution) can be described by the metric tensor
\begin{equation}
\d s^2 = -f(r)\d t^2 + \frac{\d r^2}{f(r)} +r^{2}\d{\Omega}^2_{d-2},
\end{equation}
where $\d{\Omega}^2_{d-2}$ is the canonical round metric of a unit $(d-2)$-sphere, and the function $f(r)$ takes the form
\begin{equation}
f(r) = 1-\frac{2M}{r^{d-3}} + \frac{Q^2}{r^{2d-6}}.
\end{equation}
Here $M$ and $Q$ are related to the ADM mass and electric charge of the black hole (they differ from the ADM counterparts by some constant factors, the exact forms of which are not important in this work). Note that $M$ and $Q$ have different physical dimensions for different $d$.
Without loss of generality we assume that $Q>0$.
The black hole radius $r_+$ satisfies $f(r_+)=0$ so we have
\begin{equation}
r_+=\big(M+\sqrt{M^2-Q^2}\big)^{\frac{1}{d-3}}.
\end{equation}

The temperature is proportional to the surface gravity at the black hole horizon, and is explicitly given by
\begin{equation}
T=\frac{(d-3)\left(M^2-Q^2+M\sqrt{M^2-Q^2}\right)\hbar}{2\pi\left(M+\sqrt{M^2-Q^2}\right)^{\frac{2d-5}{d-3}}}.
\label{temperature}
\end{equation}

Now we give a brief discussion on the Stefan-Boltzmann law in $(d-1)$-dimensional space (spatial dimension only). See also \cite{Vos1989,0510002}.
Considering a $(d-1)$-dimensional cavity, the number of modes $N$ is given by
\begin{equation}
\d N=\frac{\d x_1 \d x_2\ldots \d x_{d-1}\d p_1 \d p_2\ldots \d p_{d-1}}{h^{d-1}},
\end{equation}
where $h$ is the Planck constant.
Integrating the above formula and multiplying the result by a factor of $(d-2)$ because of the $(d-2)$ independent polarizations of radiation (see e.g. \cite{0510002,Lapidus1982,0011070} for more discussion in this field), we obtain
\begin{equation}
V\frac{(d-2)Ap^{d-2}\d p}{h^{d-1}}=V\frac{(d-2)A\omega^{d-2}\d\omega}{(2\pi)^{d-1}},
\end{equation}
where
\begin{equation}
A=\frac{(d-1)\pi^{\frac{d-1}{2}}}{\Gamma(\frac{d+1}{2})}
\end{equation}
is the area of an $(d-1)$-dimensional sphere with unit radius and $V$ is the volume of the $(d-1)$-dimensional cavity. For example, in 4-dimensions, we have $d=4$ and $A=4\pi$. After multiplying by the photon energy $\hbar\omega$ and by the Bose-Einstein factor
$1/(e^{\frac{\hbar\omega}{T}}-1)$, we obtain the $(d-1)$-dimensional photon gas energy
\begin{eqnarray}
V\frac{(d-2)A}{(2\pi)^{d-1}}\int \hbar \omega\frac{\omega^{d-2}\d\omega}{e^{\frac{\hbar\omega}{T}}-1}
=V\frac{(d-2)A T^{d}}{(2\pi)^{d-1}\hbar^{d-1}}\int\frac{x^{d-1}\d x}{e^x-1},
\end{eqnarray}
where the integration domain is from $0$ to $\infty$. We can find the energy density is proportional to $T^d$ and the Stefan-Boltzmann constant reads
\begin{equation}
a_d=\frac{(d-2)A}{(2\pi)^{d-1}\hbar^{d-1}}\int\frac{x^{d-1}\d x}{e^x-1}.
\end{equation}
For $d=4$, we have the well-known result for the total energy density
\begin{equation}
U=\frac{\pi^2}{15\hbar^3}VT^4.
\end{equation}

Next we need to calculate the cross section of the effective emitting surface of the black hole. We shall assume that the emitted massless particle moves along null geodesics and following \cite{HW,1904.06503,Frolov2011}, we shall apply the geometrical optics approximation. Orienting the extra $(d-3)$ angular coordinates in $\mathrm{d}\Omega_{d-2}^{2}$ and normalizing the affine parameter $\lambda$, the geodesic equation of the massless quanta reads
\begin{align}
\bigg(\frac{\mathrm{d}r}{\mathrm{d}\lambda}\bigg)^2=E^2-J^2\frac{f(r)}{r^2},
\end{align}
where $E=f(r)\frac{\mathrm{d}t}{\mathrm{d}\lambda}$ and $J=r^2\frac{\mathrm{d}\theta}{\mathrm{d}\lambda}$ are the energy and angular momentum respectively.
For the emitted massless particle to reach infinity rather than falling back to the black hole horizon, defining $b\equiv {J}/{E}$, one must have
\begin{align}
\frac{1}{b^2}\geqslant \frac{f(r)}{r^2}
\end{align}
for any $r\geqslant r_+$. Therefore we need to find the maximal value of ${f(r)}/{r^2}$. In our system it corresponds to the unstable photon orbit $r_p$, and the impact factor $b_c$ can be defined as $b_c\equiv {r_p}/{\sqrt{f(r_p)}}$. The exact formula for $r_p$ and $b_c$ are quite complicated and we omit them here. The cross section of the black hole is just the volume of $(d-2)$-dimensional sphere with radius $b_c$, that is,
\begin{align}
\sigma=\frac{\pi^{\frac{d}{2}-1}}{\Gamma(\frac{d}{2})} b_c^{d-2}.
\end{align}

Finally we obtain the mass loss rate of the black hole due to emission of massless particles via the simple differential equation
\begin{align}
\frac{\mathrm{d} M}{\mathrm{d}t}=-a_d \sigma T^d.
\label{massloss1}
\end{align}
This is the generalization of the four-dimensional Stefan-Boltzmann law in $d$-dimensional spacetime. Note that the Stefan-Boltzmann constant depends on spacetime dimension. With this understanding, we shall hereinafter write $a \equiv a_d$ for simplicity. The greybody factor is ignored because we are not interested in the exact lifetime of the evolution but rather its qualitative behaviors.

The charge-loss rate of the black hole, on the other hand, is described in the framework of quantum electrodynamics (QED). We can study particle creation from the vacuum by external electric fields in arbitrary spacetime dimensions. If the radius of black hole is much larger than the Compton wavelength of the electron, the production of electron-positron pairs is well described by ordinary flat-space QED. The rate of pair creation per unit $d$-volume reads \cite{9603152}
\begin{equation}
\Gamma=\frac{2^{\left\lfloor d/2 \right\rfloor}-1}{(2\pi)^{d-1}}\bigg(\frac{m}{\hbar}\bigg)^d\bigg(\frac{E}{E_c}\bigg)^{d/2}\mbox{exp}\bigg\{\frac{-\pi E_c}{E}\bigg\},
\end{equation}
where $E_c\equiv m^2/\hbar e$ represents the Schwinger critical charge, and $\left\lfloor~\right\rfloor$ denotes the floor function. For the charged black hole in $d$-dimensional spacetime $(d \geqslant 4)$, the electric field\footnote{The constant prefactor of the electric field depends on the definition of the charge and mass parameters, which differ from the ADM quantities by constant factors. The electric field is computed from the first law of black hole thermodynamics, since $\d M = \Phi \d Q$ gives the electric potential, from which we can compute the electric field. The fact that $E$ seems to vanish in $(2+1)$-dimensions is not alarming since Reissner-Nordstr\"om metric is not a solution in $d=3$, so we do not expect the prefactor computed this way to correctly match the result obtained directly via Gauss law of point charge for the $d=3$ case.} is given by
\begin{equation}
E=(d-3)Q/r^{d-2}.
\end{equation}

The charge loss rate of the black hole is obtained by integrating the above formula over the entire space outside the horizon. We have
\begin{align}
\frac{\d Q}{\d t}=&eA\frac{2^{\left\lfloor\frac{d}{2}\right\rfloor-1}}{(2\pi)^{d-1}}\int^{\infty}_{r_+}\bigg(\frac{(d-3)Qe}{\hbar r^{d-2}}\bigg)^{\frac{d}{2}}r^{d-2} \\ \notag
&\times \mbox{exp}\bigg\{\frac{-\pi m^2 r^{d-2}}{(d-3)Q\hbar e}\bigg\}\d r.
\end{align}
The integral can be found explicitly, although this requires somewhat tedious calculations. Unlike the 4-dimensional case, we need to make use of the properties of a special function called the ``WhittakerM function'',
\begin{align}
\mbox{WhittakerM}\left(a-\frac{1}{2},a,z\right)=2ae^{z/2}z^{\frac{1}{2}-a}[\Gamma(2a)-\Gamma(2a,z)]
\end{align}
and expand the incomplete Gamma function $\Gamma(2a,z)$ in large $z$,
\begin{equation}
\Gamma(2a,z)=z^{2a-1}e^{-z}\sum_{k=0}^{n-1}\frac{(2a-1)(2a-2)\cdots (2a-k)}{z^k}.
\end{equation}
Finally, we obtain the leading order term of the differential equation that governs charge loss, which reads
\begin{align}\label{chargeloss}
\frac{\d Q}{\d t} \approx &-\frac{eA2^{\left\lfloor\frac{d}{2}\right\rfloor-1}}{(2\pi)^{d-1}}\frac{\left( (d-3)Qe\right)^{\frac{d}{2}+1}}{\pi \hbar^{\frac{d}{2}-1}m^2}\frac{r_+^{-\frac{d^2}{2}+d+1}}{d-2} \\ \notag
&\times \mbox{exp}\left\{\frac{-\pi m^2 r_+^{d-2}}{(d-3)Q\hbar e}\right\}.
\end{align}
This is a \emph{universal result} in arbitrary dimensions $d \geqslant 4$. For $d=4$, the WhittakerM function can be described by the error function erf($r$), and the final result reads
\begin{equation}\label{chargeloss0}
\frac{\d Q}{\d t} \approx - \frac{e^4}{2\pi^{3} \hbar m^2} \frac{Q^3}{r_+^3} \exp{\left(\frac{-\pi m^2 r_+^{2}}{Q\hbar e}\right)}.
\end{equation}
This is exactly the same result obtained by HW and it is also consistent with the result using the semiclassical method to investigate the tunneling of particles through the effective Dirac gap outside of the black hole \cite{Khriplovich}.

The created charged particle will be repelled towards infinity.
Note that since we have assumed $Q>0$, it is actually the \emph{positrons}  that are emitted in this manner (because like charges repel). Similar discussion holds also for electrons, and so for simplicity, we shall use the term ``electron'' in the following discussions. (Since we are more used to refer to a particle rather than its anti-particle.)
Since the rest mass of the electron or positron is much smaller than its electrical charge, the potential energy gained by the repelled particle, $Qe/r_+^{d-3}$, is also much larger than its rest mass. Consequently, the black hole mass loss rate due to the electromagnetic pair creation is
\begin{equation}
\frac{\d M}{\d t}=\frac{Q}{r_+^{d-3}}\frac{\d Q}{\d t}.
\label{massloss2}
\end{equation}
This is also consistent with the first law of black hole thermodynamics.

Combining Eq.\eqref{massloss1} and Eq.\eqref{massloss2},  we  finally obtain the total rate of mass loss for the charged evaporating
black hole in $d$-dimensional spacetime:
\begin{equation}
\frac{\d M}{\d t}=-a \sigma T^d+\frac{Q}{r_+^{d-3}}\frac{\d Q}{\d t}.
\label{massloss}
\end{equation}
This of course recovers the model of Hiscock and Weems when $d=4$.
Note that Eq.(\ref{massloss}) forms a set of coupled ordinary differential equations with the charge loss rate given by Eq.\eqref{chargeloss}.

A few remarks are in order before we proceed. First of all, let us explain that in this model, the black holes are required to be sufficiently large. This is because the Hawking temperature is inversely proportional to the black hole size. For small black holes, the temperature is large enough to effectively create charged particles as well, in which case it would not make sense to separate the mass loss due to massless charge-less particle and massive charged particles in the way it is done in Eq.(\ref{massloss}). Only in the large mass regime one can model Hawking evaporation via HW model. See also \cite{gibbons}. Secondly, in the series expansion of the charge loss formula, we considered only the leading order term, this is amount to ignoring other charged particles which are heavier than the electrons (since the heavier a particle is, its production rate will be suppressed even more).

\section{Evaporation of Higher Dimensional Reissner-Nordstr\"om Black Hole}

Since we have obtained the coupled set of differential equations, Eq.(\ref{massloss}), which can be used to describe the evaporative evolution of the charged black hole in $d$-dimensional spacetime, we can investigate the evolution process by numerically integrating the differential equations. Before we do that, we give a brief discussion on the validity of this method. As we emphasized earlier, the production of electron-positron pairs is well described by ordinary flat-space QED if the radius of black hole is much larger than the Compton wavelength of the electron. This requires $r_+\gg {\hbar}/{m}$. Furthermore, the electric field of the black hole must be much smaller than the Schwinger limit $E_c$, otherwise the electric field is expected to become nonlinear and the black hole will
discharge rapidly. The electric field of the black hole reads
\begin{align}
E=\frac{(d-3)Q}{r_+^{d-2}}=\frac{(d-3)Q}{\left(M+\sqrt{M^2-Q^2}\right)^{\frac{d-2}{d-3}}}.
\end{align}
For a given black hole mass $M$, the electric field is a monotonic function of $Q$ and it attains maximal value at the extremal limit $Q=M$, which reads
\begin{align}
E_{\text{ext}}=\frac{d-3}{M^{\frac{1}{d-3}}}.
\end{align}
Thus in order to let the electric field of the black hole be much smaller than the Schwinger limit $E_c$, we must have
\begin{align}
M\gg \left(\frac{d-3}{E_c}\right)^{d-3}.
\end{align}
For $d=4$, the condition is simply $M\gg {1}/{E_c}$. However, for the black holes in higher dimensions, their mass must be some powers of the inverse of Schwinger critical field so that they are large enough to retain a significant charge for more than a microscopic duration.

Now we are ready to investigate the mass and charge loss of these black holes. As we have discussed, in Hiscock and Weems model, the mass loss of the black hole comes from two terms: the term $-a \sigma T^d$, which corresponds to the thermal radiation (Stefan-Boltzmann law), which depends on powers of  $M$ and $Q$, and the term $({Q}/{r_+^{d-3}}){\d Q}/{\d t}$, which corresponds to the charge loss rate multiplied by the potential energy of the electric field, which depend exponentially on $M$ and $Q$.

\begin{figure}
\begin{center}
\includegraphics[width=0.45\textwidth]{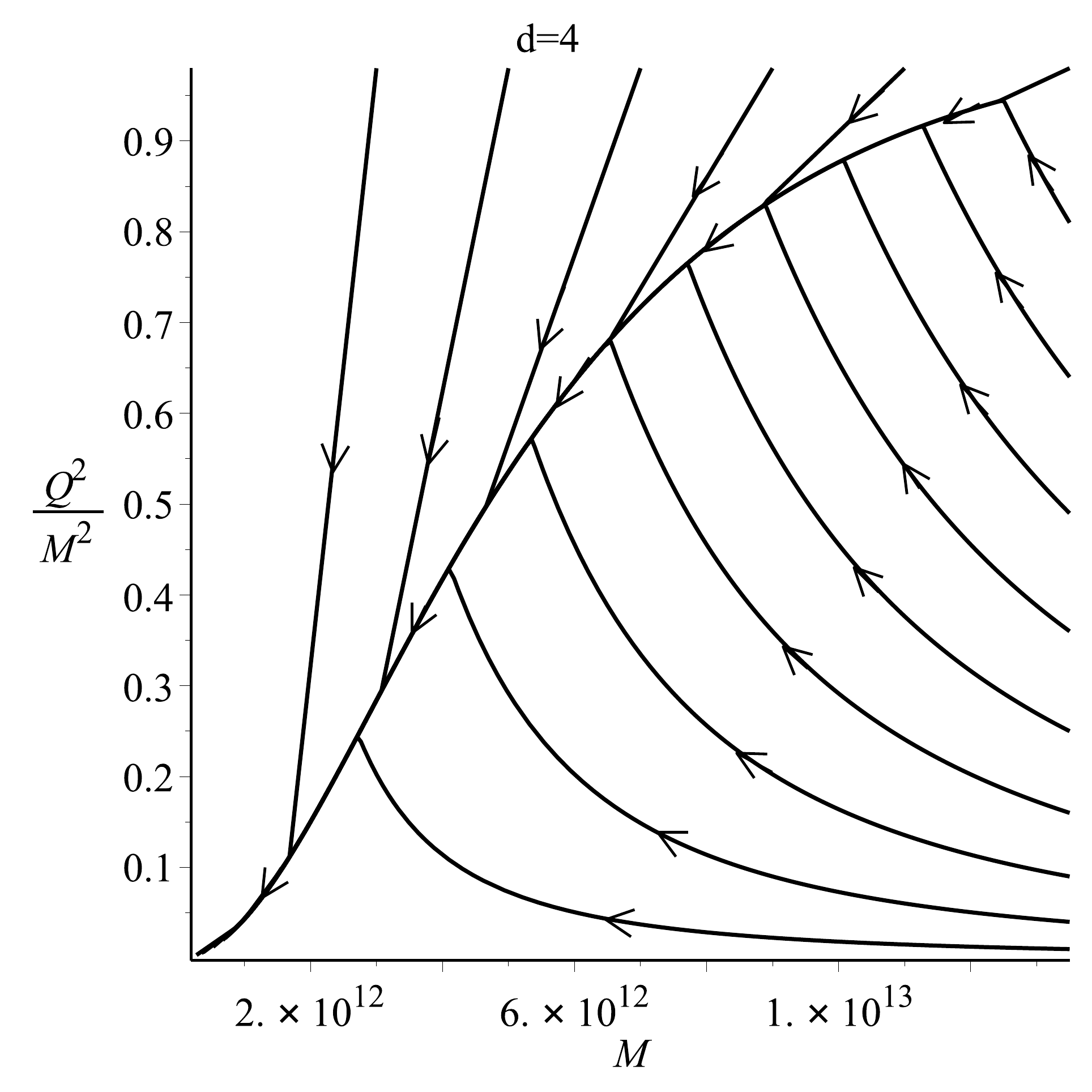}
\caption{Evaporative evolution of the 4-dimensional asymptotically flat Reissner-Nordstr\"om black holes in $\frac{Q^2}{M^2}$-$M$ diagram.}
\label{fig1}
\end{center}
\end{figure}

\begin{figure}
\begin{center}
\includegraphics[width=0.45\textwidth]{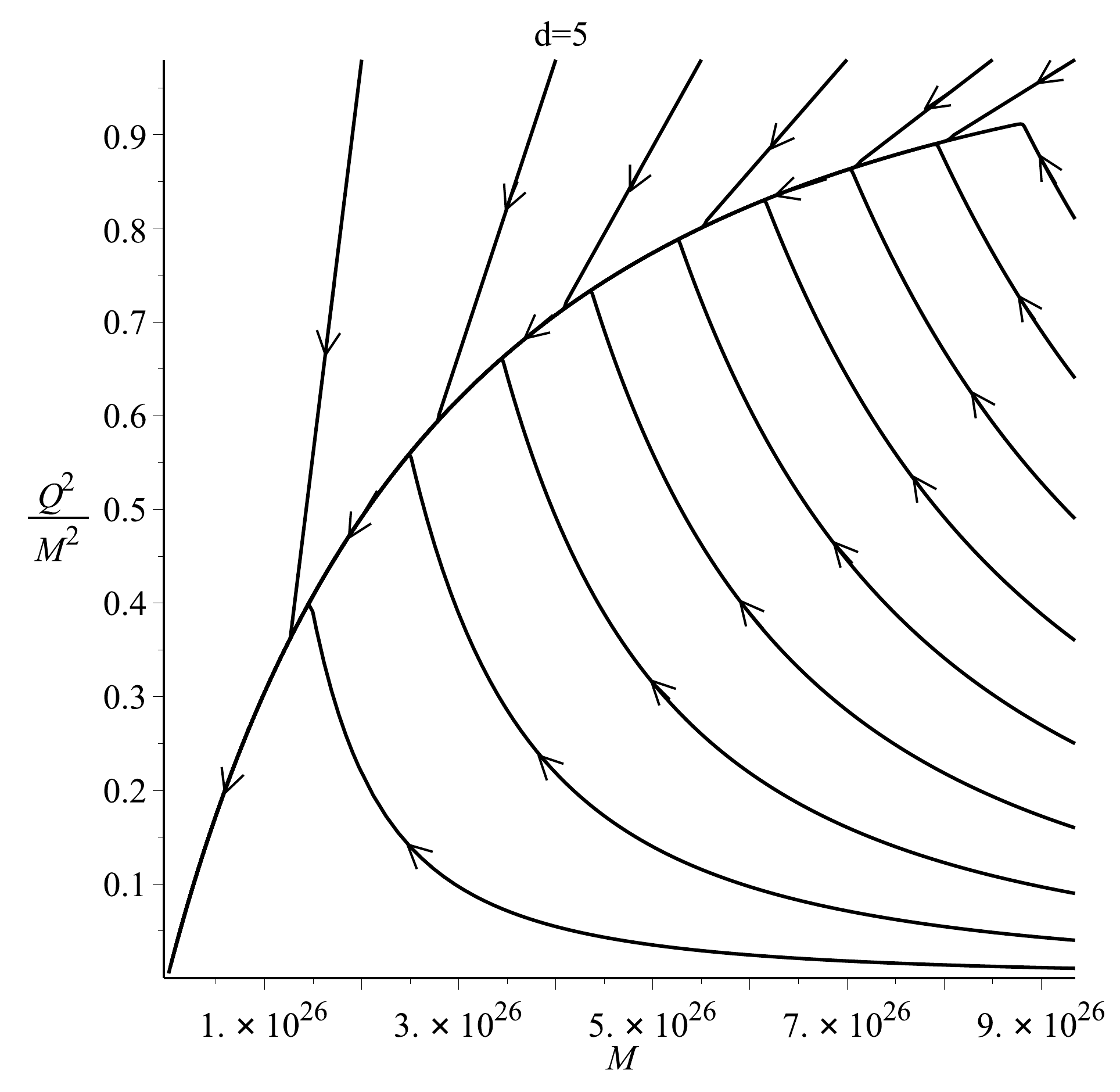}
\caption{Evaporative evolution of the 5-dimensional  asymptotically flat Reissner-Nordstr\"om black holes in $\frac{Q^2}{M^2}$-$M$ diagram.}
\label{fig2}
\end{center}
\end{figure}

\begin{figure}
\begin{center}
\includegraphics[width=0.45\textwidth]{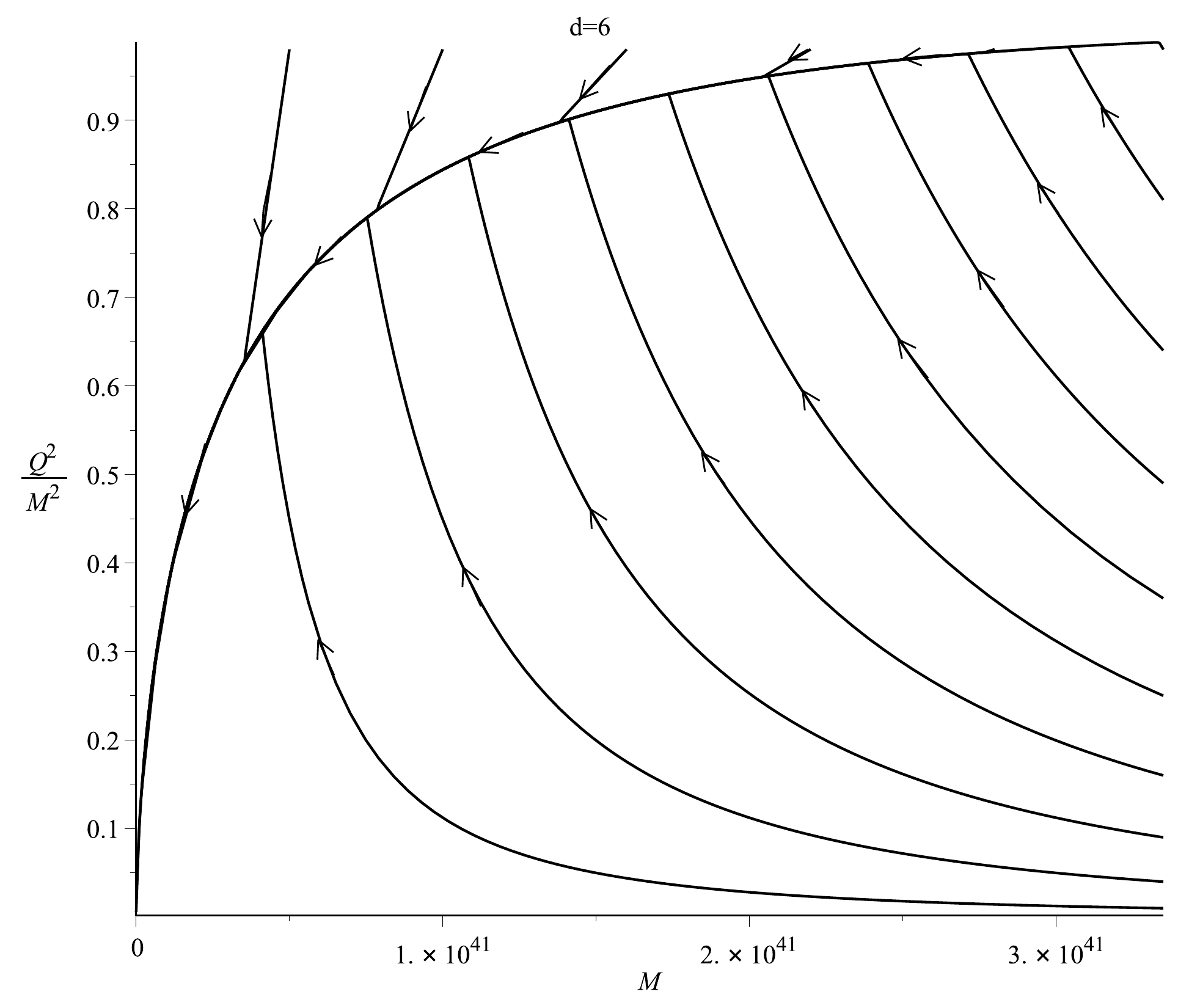}
\caption{Evaporative evolution of the 6-dimensional  asymptotically flat Reissner-Nordstr\"om black holes in $\frac{Q^2}{M^2}$-$M$ diagram.}
\label{fig3}
\end{center}
\end{figure}

For a given initial mass satisfying $M_0 \gg {1}/{E_c^{d-3}}$, if the initial charge $Q_0$ is small, the charge loss rate is suppressed due to the exponent, thus it is in the ``mass dissipation zone''. The black hole loses its mass primarily by the thermal radiation, until the electric field becomes larger and the charged loss is significant (though still suppressed by the Boltzmann factor). Then the black hole evolution curve joins the ``attractor'' in the $\frac{Q^2}{M^2}$-$M$ diagram, which is a very narrow region containing many different paths. Once on the attractor, the black hole will continue to lose charge and mass, and eventually evolves towards a Schwarzschild state. If the initial charge $Q_0$ is large, the temperature $T$ becomes small and the black hole loses its mass primarily via charge loss, until it joins the attractor.
In Fig.\eqref{fig1}, \eqref{fig2}, \eqref{fig3} we present some examples of the evaporative evolution of  charged black holes  for 4, 5,  and 6 dimensional spacetimes, respectively.

Therefore, from these plots we can conclude that the qualitative features for evaporative evolution of the charged black hole are the same in higher dimensions.
Although we have only plotted up to $d=6$ dimensions, similar results probably also hold in higher dimensions since from our analysis, it is clear that the Stefan-Boltzmann law and the Schwinger effect in different dimensions are qualitatively similar.

Note that for a fixed value of the charge-to-mass ratio $Q^*/M^*$ close to unity, it requires that a black hole that starts from the mass dissipation zone to have higher initial mass in order to reach it.
The caveat is that, in making our comparisons, we have assumed that the values of the electrical charge and mass of the electron to be the same in different dimensions (the physical dimensions of charge and mass are, however, not the same in different dimensions). We are concerned with how the results in different dimensions compare, \emph{if everything else being equal}.

\section{Conclusion: Hawking Evaporation Obeys Cosmic Censorship}

In this work we have generalized the Hiscock and Weems model of evaporating charged black holes to higher dimensions. At the technical level, we see that it involves the WhittakerM function, whereas in $d=4$ it only requires the error function.
By solving the coupled differential equations of the mass loss and charge loss rate numerically,
we have demonstrated that Reissner-Nordstr\"om black holes in higher dimensions (with explicit calculations in $d=5,6$) behave in qualitatively the same way as 4-dimensional case. This means that these black holes cannot achieve extremality via Hawking evaporation, which in turn implies that they would not evolve into naked singularities.

In the Introduction, we mentioned that the role of spacetime dimension in determining whether weak cosmic censorship is easier or harder to be violated is non-trivial. If one considers black hole collisions, then it is easier to violate cosmic censorship in higher dimensions \cite{1812.05017}, and in view of a variety of higher dimensional cases such as black strings that form naked singularities under Gregory-Laflamme instability, it would seem to suggest that it is harder to maintain cosmic censorship in higher dimensions. Nevertheless, at this stage there is no general proof for this claim, and one needs to check it for different physical mechanism, especially in view of the rather complicated situation in the case of the violation of the strong cosmic censorship we discussed in the Introduction.

Our work suggests that as the number of spacetime dimension grows, the initial mass of the black hole has to be larger in order for it to evolve towards a near extremal $Q/M$. In this sense then, one could say that as far as Hawking radiation is concerned, cosmic censorship in Reissner-Nordstr\"om spacetime is safest in $d=4$. Whether this remains to be the case in other black hole spacetimes would require more studies in the future.

\begin{acknowledgments}
 Hao Xu and Man-Hong Yung acknowledge the support by the Key-Area Research and Development Program of Guangdong Province (Grant No. 2018B030326001), the National Natural Science Foundation of China (Grant No.11875160 and No.U1801661), the Natural Science Foundation of Guangdong Province (Grant No.2017B030308003), the Guangdong Innovative and Entrepreneurial Research Team Program (Grant No. 2016ZT06D348), the Science, Technology and Innovation Commission of Shenzhen Municipality (Grant No.JCYJ20170412152620376, No.JCYJ20170817105046702, and No.KYTDPT20181011104202253), the Economy, Trade and Information Commission of Shenzhen Municipality (Grant No.201901161512). Yen Chin Ong thanks the National Natural Science Foundation of China (No.11705162) and the Natural Science Foundation of Jiangsu Province (No.BK20170479) for funding support.
\end{acknowledgments}

\end{document}